\documentclass[aps,prd,reprint,nofootinbib,floatfix,longbibliography]{revtex4-2}
\usepackage{amsmath,amssymb,graphicx,bm}
\usepackage[T1]{fontenc}
\usepackage{orcidlink}
\begin{document}

\title{Kerr-referenced analytical parametrization of rotating black holes in dynamical Chern--Simons gravity}
\author{Sardor~Murodov\orcidlink{0000-0003-2360-4475}}
\email{s.murodov@newuu.uz}
\affiliation{New Uzbekistan University, Movarounnahr Street 1, Tashkent 100000, Uzbekistan}

\author{Bahodir~Ahmedov\orcidlink{0009-0006-7375-2731}}
\email{b.ahmedov@newuu.uz}
\affiliation{New Uzbekistan University, Movarounnahr Street 1, Tashkent 100000, Uzbekistan}

\author{Bekzod~Rahmatov\orcidlink{0009-0001-0394-650X}}
\email{rahmatovbekzod@samdu.uz}
\affiliation{Tashkent State Technical University, Tashkent 100095, Uzbekistan}

\author{Javlon Rayimbaev\orcidlink{0000-0001-9293-1838}}
\email{javlon@astrin.uz}
\affiliation{Institute of Theoretical Physics, National University of Uzbekistan, Tashkent 100174, Uzbekistan}

\author{Bobomurat~Ahmedov\orcidlink{0000-0002-1232-610X}}
\email{ahmedov@astrin.uz}
\affiliation{Institute of Theoretical Physics, National University of Uzbekistan, Tashkent 100174, Uzbekistan}
\affiliation{School of Physics, Harbin Institute of Technology, Harbin 150001, People’s Republic of China}

\date{\today}

\begin{abstract}
Rotating black holes in dynamical Chern–Simons gravity are known nonperturbatively mainly through numerical solutions, which limits their direct use in strong-field applications. We construct a Kerr-referenced analytical representation of the near-general-relativistic numerical branch using a compact radial coordinate, low-order angular multipoles, and continued fractions. The resulting two-parameter family accurately reproduces the numerical metric throughout the validated weak-coupling and moderate-spin domain. Independent off-grid solutions confirm that the parametrization retains its accuracy away from the calibration set. We further assess the model through equatorial light rings, critical impact parameters, and full two-dimensional null-geodesic ray tracing without assuming separability. The resulting shadow observables closely match those obtained from the numerical geometries. This parametrization therefore provides a practical analytical metric for geodesic and strong-field calculations without relying on a slow-rotation expansion.
\end{abstract}

\maketitle

\section{Introduction}
\label{sec:introduction}

Dynamical Chern--Simons (dCS) gravity is a parity-violating extension of
general relativity in which a pseudoscalar couples to the Pontryagin density.
The construction is rooted in the four-dimensional Chern--Simons modification
of gravity~\cite{JackiwPi2003} and is motivated by anomaly cancellation,
effective field theory, string-inspired models, and related high-energy
constructions~\cite{AlexanderYunes2009Review,AdakDereli2012}.  In its
dynamical form the scalar field evolves self-consistently, which avoids the
additional Pontryagin constraint of the nondynamical formulation
~\cite{YunesPretorius2009dCS,AlexanderYunes2009Review}.  Rotation is essential:
for a spherically symmetric vacuum black hole the Pontryagin density vanishes,
whereas a rotating geometry sources the pseudoscalar and activates the
Chern--Simons correction.  This makes rotating black holes the natural
strong-field arena for the theory.  More broadly, gravitational-wave,
X-ray, timing, and horizon-scale imaging observations provide complementary
routes to strong-gravity tests~\cite{YunesSiemens2013,ChamberlainYunes2017,
Krawczynski2012}.

Early studies of rotation in Chern--Simons gravity already showed that a
Kerr-like frame-dragging sector is the first place where parity-violating
effects become visible~\cite{KonnoMatsuyamaTanda2007,KonnoMatsuyamaTanda2009}.
For dynamical dCS gravity, Yunes and Pretorius constructed the leading
slow-rotation, small-coupling black-hole solution
~\cite{YunesPretorius2009dCS}.  Ali-Ha\"imoud and Chen developed a
complementary treatment of slowly rotating compact objects and black holes
~\cite{AliHaimoudChen2011dCS}, while Yagi, Yunes, and Tanaka extended the
geometry to quadratic order in spin, where even-parity corrections and a
modified quadrupole structure appear~\cite{YagiYunesTanaka2012dCS}.  Related
perturbative and rapidly rotating scalar-field analyses further clarified the
rotation dependence and the approach to the extremal regime
~\cite{KonnoTakahashi2014,McNeesSteinYunes2016}.  The perturbative geometries
have supported studies of geodesics, lensing, accretion disks, and wave
dynamics~\cite{AmarillaEiroaGiribet2010,ChenJing2010,HarkoKovacsLobo2010},
but their direct domain of validity is restricted by the spin and coupling
expansions.

A qualitatively different step was taken by Delsate, Herdeiro, and Radu, who
constructed stationary and axisymmetric dCS black holes by solving the
boundary-value problem directly, without expanding the stationary solution in
spin or coupling~\cite{Delsate2018ECS}.  These solutions are given numerically
in a horizon-regular Lewis--Papapetrou-type gauge and exhibit the expected
asymptotic mass, angular momentum, and scalar dipole.  They therefore provide
a natural target for an analytical reconstruction.  Here and throughout this
paper, ``nonperturbative'' refers only to the construction of the stationary
solution without a formal spin or coupling expansion.  It does not remove the
effective-field-theory limitations of dCS gravity.  The higher-derivative
structure of the full field equations and the status of the initial-value
problem motivate this distinction~\cite{DelsateHilditchWitek2015,
GarfinklePretoriusYunes2010,BertiYagiYunes2018}.  Order-reduced numerical
relativity has consequently played an important role in dynamical
applications~\cite{OkounkovaSteinScheel2017,OkounkovaSteinMoxon2020}.

The black-hole perturbation problem provides a complementary view of the same
physics.  Axial gravitational and scalar perturbations are coupled already
around Schwarzschild in dCS gravity
~\cite{CardosoGualtieri2009,MolinaPaniCardosoGualtieri2010,
MotohashiSuyama2012}.  For rotating black holes, slow-rotation perturbation
equations and quasinormal modes have been developed to increasing accuracy
~\cite{SrivastavaChenShankaranarayanan2021,WagleYunesSilva2022,
WagleLiChen2024}.  More broadly, rotating black holes in higher-derivative
effective theories have been studied through corrected Kerr geometries and
ringdown spectra~\cite{CanoRuiperez2019,CanoFransenHertog2022}, while dCS
superradiance and nonlinear scalarization illustrate additional phenomena
that can arise beyond the stationary linear-coupling sector considered here
~\cite{AlexanderGabadadzeJenks2023,DonevaYazadjiev2021}.  Closely related
higher-curvature constructions with a parity-odd source have also been explored
in Einstein--scalar--Chern--Simons gravity by introducing a NUT charge; in that
setting scalarized nutty wormholes and their domains of existence were obtained
in Refs.~\cite{IbadovKleihausKunzMurodov2021,IbadovKleihausKunzMurodov2022}.
These developments reinforce the usefulness of compact analytical metrics that
can be inserted directly into geodesic and perturbative calculations.  Recent black-hole
spectroscopy reviews emphasize the same need for accurate rotating
backgrounds when connecting modified-gravity predictions to ringdown
observables~\cite{BertiCardosoCarullo2026}.

For phenomenology, a purely numerical metric is often inconvenient because
circular orbits, ray tracing, transfer functions, accretion observables, and
perturbation equations repeatedly require the metric and its derivatives at
arbitrary points.  This motivates theory-agnostic black-hole parametrizations
such as bumpy metrics and deformed-Kerr constructions
~\cite{VigelandHughes2010,JohannsenPsaltis2011,Johannsen2013}.  A particularly
efficient alternative is the continued-fraction framework of Rezzolla and
Zhidenko~\cite{RezzollaZhidenko2014}, later extended by Konoplya, Rezzolla,
and Zhidenko to generic stationary axisymmetric metrics
~\cite{KonoplyaRezzollaZhidenko2016}.  The same framework has been tested with
ray-traced shadows~\cite{KonoplyaRezzollaZhidenko2016Shadow}, X-ray
spectroscopy~\cite{NiJiangBambi2016}, and numerical higher-curvature black
holes.  In Einstein--scalar--Gauss--Bonnet gravity, for example, low-order
continued fractions reproduce numerical metric functions and gauge-invariant
shadow observables with high accuracy~\cite{Konoplya2020EsGBCFA}.

Shadows and photon orbits are particularly useful validation observables
because they depend directly on the strong-field metric.  Chern--Simons
corrections to null geodesics and shadows were studied already in the
slow-rotation regime~\cite{AmarillaEiroaGiribet2010}, together with strong
lensing and geodetic precession~\cite{ChenJing2010}.  Related strong-field
dynamics in Chern--Simons backgrounds, including nonintegrable test-particle
motion, further illustrates the sensitivity of orbital structure to the
parity-violating sector~\cite{ZhouChenJing2021}.  More generally, particle
orbits and quasiperiodic-oscillation observables in deformed black-hole
geometries provide complementary probes of departures from the standard
Schwarzschild/Kerr picture~\cite{MurodovRayimbaevAhmedovKarimbaev2023}.
Numerical dCS shadow calculations were later performed using order-reduced
black-hole data~\cite{Okounkova2019dCSShadows}.  Fully nonlinear shadows of
higher-curvature black holes have likewise been studied in rotating
Einstein--dilaton--Gauss--Bonnet solutions~\cite{CunhaHerdeiroKleihausKunzRadu2017}.  More generally, parametrized-metric studies
and shadow reviews emphasize that observable comparisons are more informative
than component-wise metric errors alone
~\cite{RezzollaMizunoYounsi2018,PerlickTsupko2022,Berti2019KerrTests}.
This is the reason we use both equatorial light rings and full two-dimensional
ray tracing as validation tests below.

Despite the availability of the Delsate--Herdeiro--Radu numerical spinning
solutions and the broader parametrized-metric literature, our targeted
literature survey did not identify a compact Kerr-referenced
continued-fraction representation of that numerical dCS family tailored to
generic geodesic calculations.  The present work addresses that gap for the
near-GR portion of the branch.  We write each numerical geometry as an exact
Kerr background with the same $(r_H,\Omega_H)$ plus a Chern--Simons
deformation.  The radial dependence is represented by low-order continued
fractions in a compact coordinate, while the angular dependence is organized
in a small set of Legendre multipoles consistent with the equatorial parity of
the metric and pseudoscalar.  Factoring out the leading coupling and spin
scalings yields a two-parameter analytical family
$c_i(\bar\alpha,\chi_K)$ rather than a separate fit at each numerical point.
The calibrated interval is intentionally restricted to the numerical region
that passes our residual, constraint, and resolution tests; no claim is made
that the present coefficients describe the full strong-coupling DHR branch.

The final interpolation uses twelve validated numerical solutions spanning $0.005\leq\bar\alpha\leq0.010$,
and
$0.202107\leq\chi_K\leq0.549139 .$
Over this domain, the joint fit reconstructs the total metric functions with
worst relative RMS errors below $3.7\times10^{-5}$.  After freezing the CFA
coefficients, we test two completely off-grid numerical solutions whose
coupling and spin values do not coincide with the calibration grid.  The
resulting family is further validated with prograde and retrograde light rings
and critical impact parameters, followed by full two-dimensional null-geodesic
ray tracing without assuming separability.  At the strongest validated
high-spin point, the centered shadow contours agree with the numerical
geometry to a few parts in $10^5$.  Resolution convergence and the
Yunes--Pretorius perturbative benchmark provide independent checks of the
numerical implementation.

The paper is organized as follows.  Section~\ref{sec:theory} summarizes the
dCS model, boundary conditions, numerical construction, convergence tests, and
the perturbative benchmark.  Section~\ref{sec:cfa} introduces the
Kerr-referenced continued-fraction representation and the coupling--spin fit.
Section~\ref{sec:shadow_validation} validates the analytical geometry using
equatorial light rings and full two-dimensional ray tracing without assuming
separability, including the ray-tracing error budget.  The main results and
limitations are summarized in Sec.~\ref{sec:conclusions}.

\section{Theory and numerical construction}
\label{sec:theory}

\subsection{Dynamical Chern--Simons model}
\label{sec:theory_model}

We consider the massless, linearly coupled dynamical Chern--Simons (dCS)
model used for the nonperturbative stationary solutions of Delsate, Herdeiro,
and Radu~\cite{Delsate2018ECS} (where the same theory is denoted dynamical
Einstein--Chern--Simons gravity).  The action is
\begin{equation}
S=
\int d^4x\,\sqrt{-g}
\left[
\frac{R}{16\pi G}
+\frac{\alpha}{4}\,\phi\,{}^\ast RR
-\frac{1}{2}\nabla_a\phi\nabla^a\phi
\right],
\label{eq:dcs_action}
\end{equation}
where $\alpha$ has dimensions of length squared and
\begin{equation}
{}^\ast RR
=
{}^\ast R^a{}_{b}{}^{cd}R^b{}_{acd},
\qquad
{}^\ast R^{ab}{}_{cd}
=
\frac{1}{2}\epsilon_{cdef}R^{abef}
\label{eq:pontryagin}
\end{equation}
is the Pontryagin density.  The coupling is linear in the pseudoscalar and the
potential is set to zero.

The metric equations can be written as
\begin{equation}
G_{ab}
=
8\pi G
\left(
T^{(\phi)}_{ab}-2\alpha C_{ab}
\right),
\label{eq:dcs_einstein}
\end{equation}
with
\begin{equation}
T^{(\phi)}_{ab}
=
\nabla_a\phi\nabla_b\phi
-\frac{1}{2}g_{ab}
\nabla_c\phi\nabla^c\phi ,
\label{eq:scalar_stress}
\end{equation}
and
\begin{align}
C^{ab}={}&
(\nabla_c\phi)\,
\epsilon^{cde(a}\nabla_eR^{b)}{}_{d}
\nonumber\\
&+
(\nabla_c\nabla_d\phi)\,
{}^\ast R^{d(ab)c}.
\label{eq:C_tensor}
\end{align}
The scalar equation is
\begin{equation}
\Box\phi
=
-\frac{\alpha}{4}\,{}^\ast RR .
\label{eq:scalar_equation}
\end{equation}
In the numerical implementation we set $8\pi G=1$.  The metric residual used
below is therefore
\begin{equation}
E_{ab}
\equiv
G_{ab}-T^{(\phi)}_{ab}+2\alpha C_{ab}=0 .
\label{eq:rescaled_metric_eq}
\end{equation}
This convention is used consistently in all numerical and CFA coefficients
reported in the paper.

For a stationary and axisymmetric spacetime with Killing vectors
$\partial_t$ and $\partial_\varphi$, we adopt the same circular metric gauge as
Ref.~\cite{Delsate2018ECS}.  Circular stationary ansatzes are also the natural
choice for perturbatively connected stationary black holes in broad classes of
gravitational effective theories~\cite{XieZhangSilva2021}.  The line element is
\begin{align}
ds^2={}&
-e^{2F_0}N\,dt^2
+e^{2F_1}
\left(
\frac{dr^2}{N}+r^2d\theta^2
\right)
\nonumber\\
&+
e^{2F_2}r^2\sin^2\theta
(d\varphi-Wdt)^2,
\qquad
N=1-\frac{r_H}{r},
\label{eq:numerical_metric_ansatz}
\end{align}
where $F_0,F_1,F_2,W$, and $\phi$ depend only on $(r,\theta)$.
The horizon is located at $r=r_H$.

\subsection{Boundary conditions and asymptotic charges}
\label{sec:boundary_conditions}

Asymptotic flatness requires
\begin{equation}
F_0=F_1=F_2=W=\phi=0,
\qquad r\rightarrow\infty .
\label{eq:bc_infinity}
\end{equation}
On the symmetry axis,
\begin{equation}
\partial_\theta F_i
=
\partial_\theta W
=
\partial_\theta\phi
=0,
\qquad
F_1=F_2,
\qquad \theta=0,
\label{eq:bc_axis}
\end{equation}
where the last relation removes a conical singularity.  Equatorial reflection
symmetry gives
\begin{equation}
\partial_\theta F_i
=
\partial_\theta W
=0,
\qquad
\phi=0,
\qquad
\theta=\frac{\pi}{2}.
\label{eq:bc_equator}
\end{equation}
Thus the metric functions are equator-even whereas the pseudoscalar is
equator-odd.

Following Ref.~\cite{Delsate2018ECS}, we introduce the horizon-regular radial
coordinate
\begin{equation}
x=\sqrt{r^2-r_H^2}.
\label{eq:x_coordinate}
\end{equation}
At $x=0$ the boundary conditions are
\begin{equation}
\partial_xF_i=0,
\qquad
\partial_x\phi=0,
\qquad
W=\Omega_H .
\label{eq:bc_horizon}
\end{equation}
The leading asymptotic behavior takes the form
\begin{align}
F_0&=\frac{c_t}{r}+O(r^{-2}),
&
F_1&=-\frac{c_t}{r}+O(r^{-2}),
\nonumber\\
F_2&=-\frac{c_t}{r}+O(r^{-2}),
&
W&=\frac{c_\varphi}{r^3}+O(r^{-4}),
\nonumber\\
\phi&=\frac{q\cos\theta}{r^2}+O(r^{-3}),
\label{eq:asymptotic_fields}
\end{align}
so that
\begin{equation}
M=\frac{r_H-2c_t}{2},
\qquad
J=\frac{c_\varphi}{2}.
\label{eq:ADM_charges}
\end{equation}
The scalar has no monopole term; $q$ is its dipole moment.

The equations possess the scale transformation
\begin{align}
r&\rightarrow\lambda r,
& r_H&\rightarrow\lambda r_H,
& \alpha&\rightarrow\lambda^2\alpha,
\nonumber\\
M&\rightarrow\lambda M,
& J&\rightarrow\lambda^2J .
\label{eq:scaling_symmetry}
\end{align}
We therefore set $r_H=1$ in the numerical construction and restore the scale
through
\begin{equation}
\bar\alpha=\frac{\alpha}{r_H^2}.
\label{eq:alpha_bar_def}
\end{equation}

For comparison with Ref.~\cite{Delsate2018ECS}, it is also useful to quote
their dimensionless coupling
\begin{equation}
\xi=\frac{\alpha\sqrt{8\pi G}}{M^2}.
\label{eq:DHR_xi}
\end{equation}
After the field rescaling and unit choice $8\pi G=1$ used in our numerical
implementation, this becomes $\xi=\alpha/M^2$.  The common CFA calibration
rectangle used below therefore corresponds to approximately
\begin{equation}
0.014\lesssim \xi \lesssim 0.038 .
\label{eq:xi_calibration_range}
\end{equation}
Thus, although the stationary equations are solved without a perturbative
expansion in $\alpha$ or spin, the present calibrated family samples the
near-GR, weak-coupling portion of the broader numerical dCS branch.  This
distinction is important when interpreting the term ``nonperturbative'' in the
present work.

\subsection{Elliptic system and discretization}
\label{sec:numerical_scheme}

For the ansatz~\eqref{eq:numerical_metric_ansatz}, the stationary problem is
solved using the scalar equation together with four combinations of the metric
equations,
\begin{align}
\Box\phi+\frac{\alpha}{4}{}^\ast RR&=0,
\nonumber\\
E^x{}_{x}+E^\theta{}_{\theta}&=0,
& E^\varphi{}_{\varphi}&=0,
\nonumber\\
E^t{}_{t}&=0,
& E^t{}_{\varphi}&=0.
\label{eq:elliptic_system}
\end{align}
The remaining equations,
\begin{equation}
{\cal C}_1\equiv E^x{}_{\theta}=0,
\qquad
{\cal C}_2\equiv
E^x{}_{x}-E^\theta{}_{\theta}=0,
\label{eq:monitoring_constraints}
\end{equation}
are not imposed as independent equations and are instead used as numerical
constraints.  This follows the equation split used in
Ref.~\cite{Delsate2018ECS}, where the same two residual combinations are
monitored independently of the five elliptic equations.

The third derivatives entering the Chern--Simons tensor are reduced by
introducing derivative fields
\begin{align}
S_1&=\partial_rF_0,
& S_2&=\partial_rF_2,
& S_3&=\partial_rW,
\nonumber\\
Q_1&=\partial_\theta F_0,
& Q_2&=\partial_\theta F_2,
& Q_3&=\partial_\theta W .
\label{eq:auxiliary_fields}
\end{align}
In our production solver these quantities are reconstructed from the primary
fields at each nonlinear iteration and inserted into the reduced
Chern--Simons tensor.  This avoids applying a third numerical derivative
directly to the primary metric functions.

We compactify the radial domain using
\begin{equation}
\rho=\frac{x}{r_H},
\qquad
u=\frac{\rho}{1+\rho},
\label{eq:numerical_compactification}
\end{equation}
such that $u=0$ is the horizon and $u=1$ is spatial infinity.  Equatorial
parity allows us to solve only
\begin{equation}
0\le u\le1,
\qquad
0\le\theta\le\frac{\pi}{2}.
\label{eq:numerical_domain}
\end{equation}
The final wide-spin data set is computed on an equidistant
$24\times16$ grid using five-point Fornberg finite-difference stencils,
which are fourth order in the bulk.  The exact Kerr solution with the desired
$(r_H,\Omega_H)$ provides the $\alpha=0$ seed, and the dCS branch is generated
by continuation in $\alpha$.

A direct finite-difference evaluation of the curvature tensors has a small
but non-negligible truncation floor even when evaluated on exact Kerr.  We
suppress this floor with a Kerr control variate.  Denoting a discrete operator
by a subscript $h$, we use
\begin{align}
\widehat G_h[g]
&=
G_h[g]-G_h[g_K],
\label{eq:cv_G}\\
\widehat{\cal P}_h[g]
&=
{\cal P}_h[g]-{\cal P}_h[g_K]
+{\cal P}^{\rm an}_K,
\label{eq:cv_P}\\
\widehat C_h[g,\phi]
&=
C_h[g,\phi]-C_h[g_K,\phi]
+C^{\rm vac}_K[\phi],
\label{eq:cv_C}
\end{align}
where ${\cal P}\equiv{}^\ast RR$,
${\cal P}^{\rm an}_K$ is the analytical Kerr Pontryagin density, and
$C^{\rm vac}_K$ is evaluated with the exact vacuum Ricci tensor.  Since exact
Kerr obeys $G_{ab}=R_{ab}=0$, this replacement removes most of the discrete
Kerr background error without changing the continuum equations.

For a fixed metric we solve the scalar equation with a sparse linear solver.
The four metric equations are then updated by a Newton--Krylov step.  The
finite-difference Jacobian is assembled with graph coloring and solved with an
ILU-preconditioned GMRES iteration.  The scalar and Chern--Simons sources are
then refreshed and the process is repeated until the coupled residuals are
stationary.  This continuation/Newton procedure is substantially faster than
forming a dense Jacobian and made the two-parameter spin--coupling survey
practical.

Independent implementation checks were carried out before constructing the
solution family.  An automatic-differentiation tensor engine reproduces
Ricci-flat Kerr and the analytical Kerr Pontryagin density to essentially
machine precision.  The finite-difference auxiliary reduction was separately
checked against a direct automatic-differentiation evaluation of $C_{ab}$.

\subsection{Acceptance criteria and numerical family}
\label{sec:numerical_family}

The solved equations and the two monitoring constraints probe different
linear combinations of the field equations.  We therefore require all of
them to pass independent numerical tests.  For the metric equations we impose
\begin{equation}
\Vert R_g\Vert_2<5\times10^{-5},
\label{eq:metric_acceptance}
\end{equation}
and for the scalar equation
\begin{equation}
R_\phi^{\rm rel}<10^{-10}.
\label{eq:scalar_acceptance}
\end{equation}
The constraint diagnostics are computed away from the horizon, symmetry axis,
and compactified infinity on
\begin{equation}
0.20\le u\le0.80,
\qquad
0.15\le\theta\le\frac{\pi}{2}-0.15,
\label{eq:constraint_mask}
\end{equation}
with the requirements
\begin{equation}
{\rm RMS}({\cal C}_1)<2\times10^{-5},
\qquad
{\rm RMS}({\cal C}_2)<5\times10^{-5}.
\label{eq:constraint_acceptance}
\end{equation}
These values are internal numerical acceptance thresholds, not physical
constraints on $\alpha$.

The final rectangular calibration set contains three couplings and four
spins,
\begin{align}
\bar\alpha&=\{0.005,0.0075,0.010\},
\nonumber\\
\Omega_Hr_H&=\{0.10,0.15,0.20,0.25\}.
\label{eq:final_numerical_grid}
\end{align}
corresponding to the Kerr-reference spins listed in
Table~\ref{tab:numerical_family}.  All twelve points pass the criteria above.
The extracted $M$, $J$, and scalar dipole are also listed in the table.  The
quantity $q/\alpha$ grows monotonically with spin, as expected for a
rotation-sourced pseudoscalar.

\begin{table*}[t]
\centering
\caption{Numerical dCS solutions used to calibrate the final wide-spin CFA family. We set $r_H=1$. The reference Kerr spin $\chi_K$ is obtained from the low-spin Kerr branch with the same $(r_H,\Omega_H)$. $M$, $J$, and $q$ are extracted from the asymptotic metric and scalar-field coefficients.}
\label{tab:numerical_family}
\begin{tabular}{c c c c c c c}
\hline\hline
$\bar\alpha$ & $\Omega_H r_H$ & $\chi_K$ & $M/r_H$ & $J/r_H^2$ & $j=J/M^2$ & $q/\alpha$ \\
\hline
0.0050 & 0.10 & 0.202107 & 0.510534997 & 0.052676015 & 0.202098 & 0.125172 \\
0.0075 & 0.10 & 0.202107 & 0.510533416 & 0.052670558 & 0.202078 & 0.125208 \\
0.0100 & 0.10 & 0.202107 & 0.510531201 & 0.052662929 & 0.202051 & 0.125259 \\
0.0050 & 0.15 & 0.307646 & 0.525485854 & 0.084950881 & 0.307642 & 0.189426 \\
0.0075 & 0.15 & 0.307646 & 0.525482210 & 0.084942431 & 0.307616 & 0.189474 \\
0.0100 & 0.15 & 0.307646 & 0.525477096 & 0.084930612 & 0.307579 & 0.189542 \\
0.0050 & 0.20 & 0.420428 & 0.551075028 & 0.127675071 & 0.420421 & 0.256422 \\
0.0075 & 0.20 & 0.420428 & 0.551071016 & 0.127665173 & 0.420395 & 0.256475 \\
0.0100 & 0.20 & 0.420428 & 0.551061521 & 0.127648465 & 0.420354 & 0.256552 \\
0.0050 & 0.25 & 0.549139 & 0.598320447 & 0.196550883 & 0.549044 & 0.329596 \\
0.0075 & 0.25 & 0.549139 & 0.598309083 & 0.196534198 & 0.549018 & 0.329648 \\
0.0100 & 0.25 & 0.549139 & 0.598293096 & 0.196510845 & 0.548983 & 0.329722 \\
\hline\hline
\end{tabular}
\end{table*}

\subsection{Resolution convergence}
\label{sec:resolution_convergence}

The original DHR construction used several grids, with most published results
obtained on an equidistant $150\times30$ mesh and a Newton--Raphson solver
~\cite{Delsate2018ECS}.  Our control-variate formulation and sparse
Newton--Krylov implementation are numerically different, so the grid sizes
cannot be compared directly.  We therefore test convergence of the physical
quantities used by the CFA construction within our own discretization.

We select the common largest coupling of the final interpolation rectangle,
$\bar\alpha=0.010$, and three representative spins,
\begin{equation}
\chi_K=0.202107,\qquad 0.420428,\qquad 0.549139,
\label{eq:resolution_test_spins}
\end{equation}
corresponding to $\Omega_Hr_H=0.10$, $0.20$, and $0.25$.  Each solution is
recomputed on
\begin{equation}
24\times16\ \longrightarrow\ 32\times20\ \longrightarrow\ 40\times24
\label{eq:resolution_sequence}
\end{equation}
grids with the same five-point finite-difference stencil.  The highest-spin
$40\times24$ solution requires an additional nonlinear polishing step, after
which all three resolutions satisfy the acceptance criteria of
Eqs.~\eqref{eq:metric_acceptance}--\eqref{eq:constraint_acceptance}.

Table~\ref{tab:resolution_convergence} shows the last-step
$32\times20\rightarrow40\times24$ relative changes in the ADM mass, angular
momentum, scalar dipole, and the equatorial horizontal light-ring shadow
half-width.  The global charges $M$ and $J$ and the geodesic observable $R_h$
are stable at approximately the $10^{-5}$ level.  The scalar dipole converges
more slowly, with a last-step change of about $1.1$--$1.2\times10^{-3}$.
This slower convergence is consistent with the fact that $q$ is extracted
from the subleading asymptotic scalar tail rather than from the strong-field
metric itself.

\begin{table}[t]
\centering
\caption{Resolution convergence at $\bar\alpha=0.010$. We quote $\delta X=|X_{40\times24}-X_{32\times20}|/|X_{40\times24}|$ and the $40\times24$ value of the unsolved constraint ${\cal C}_2$.}
\label{tab:resolution_convergence}
\resizebox{\columnwidth}{!}{%
\begin{tabular}{c c c c c c}
\hline\hline
$\chi_K$ & $\delta M$ & $\delta J$ & $\delta q$ & $\delta R_h$ & ${\rm RMS}({\cal C}_2)$ \\
\hline
0.202107 & $4.57\times10^{-6}$ & $1.03\times10^{-5}$ & $1.12\times10^{-3}$ & $7.12\times10^{-6}$ & $1.25\times10^{-5}$ \\
0.420428 & $5.16\times10^{-6}$ & $1.66\times10^{-5}$ & $1.18\times10^{-3}$ & $1.22\times10^{-5}$ & $2.97\times10^{-5}$ \\
0.549139 & $9.58\times10^{-6}$ & $4.09\times10^{-7}$ & $1.24\times10^{-3}$ & $1.17\times10^{-5}$ & $3.49\times10^{-5}$ \\
\hline\hline
\end{tabular}}
\end{table}

The nonlinear residual norms are not strictly monotonic with grid size because
each resolution is solved to a finite nonlinear stopping tolerance rather than
by evaluating a single fixed continuum field on successively finer meshes.
The independently monitored constraint ${\cal C}_2$, however, remains below
the adopted threshold at every resolution, and the extracted global and
geodesic quantities remain stable as summarized above.  We therefore use the
$24\times16$ grid for the full two-parameter CFA calibration, while the
higher-resolution sequence provides an independent estimate of the numerical
uncertainty of the observables entering the fit.

\subsection{Perturbative external benchmark}
\label{sec:YP_benchmark}

As an external validation of the normalization and scalar sector, we compare
our numerical solutions with the slow-rotation, weak-coupling result of
Yunes and Pretorius~\cite{YunesPretorius2009dCS}.  In the canonical scalar
normalization used by our rescaled equations, their leading solution reads
\begin{equation}
\phi_{\rm YP}
=
\frac{5}{8}\,
\alpha\chi_K\,
\frac{\cos\theta}{R^2}
\left(
1+\frac{2M_K}{R}
+\frac{18M_K^2}{5R^2}
\right),
\label{eq:YP_scalar}
\end{equation}
where $R$ is the Boyer--Lindquist radius of the Kerr reference geometry.
Consequently, the asymptotic scalar dipole must satisfy
\begin{equation}
\frac{q}{\alpha\chi_K}
\longrightarrow
\frac{5}{8}
\label{eq:YP_dipole_limit}
\end{equation}
in the joint weak-coupling and slow-spin limit.

We test this prediction at $\alpha=0.005$ and
$\Omega_Hr_H=0.025$, for which the Kerr-reference spin is
$\chi_K=0.0500313$.  The extracted dipole coefficient converges as
\begin{equation}
\frac{q}{\alpha\chi_K}
=
0.621820,\quad
0.623348,\quad
0.624000
\label{eq:YP_resolution_sequence}
\end{equation}
on the $24\times16$, $32\times20$, and $40\times24$ grids,
respectively.  The finest-grid value differs from the analytical
$5/8=0.625$ coefficient by only $1.6\times10^{-3}$ in relative terms.

\begin{figure}[t]
\centering
\includegraphics[width=0.96\columnwidth]{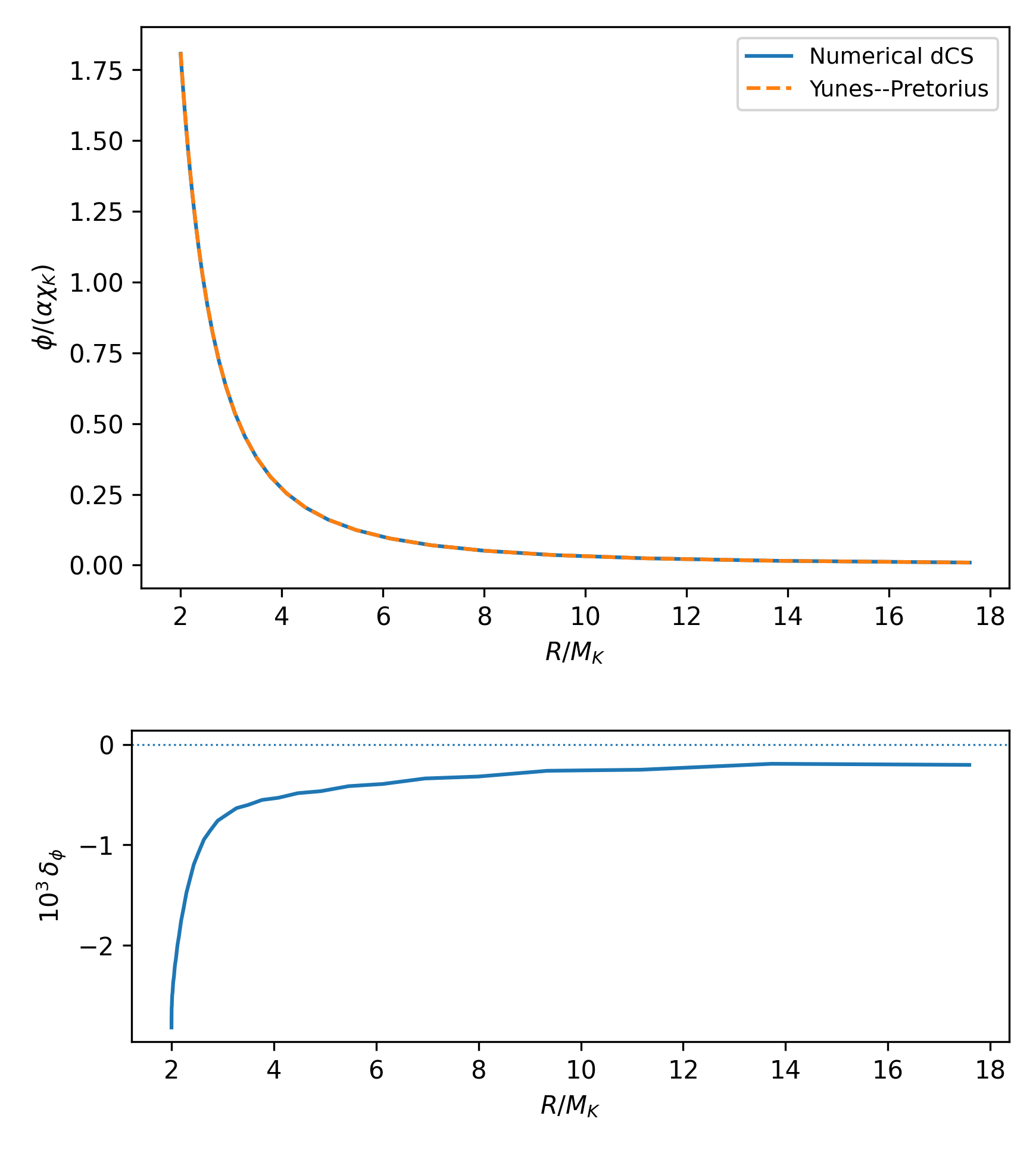}
\caption{External perturbative benchmark at
$\alpha=0.005$ and $\chi_K=0.0500313$.  The upper panel compares the
numerical dCS scalar profile on the rotation axis with the leading
Yunes--Pretorius slow-rotation, weak-coupling solution.  Because the two
profiles nearly overlap at the scale of the main panel, the lower panel shows
$10^3\delta_\phi$, where
$\delta_\phi=(\phi_{\rm num}-\phi_{\rm YP})/\phi_{\rm YP}$.}
\label{fig:YP_scalar_benchmark}
\end{figure}

The comparison is not restricted to the asymptotic coefficient.  Using the
exact Kerr radial transformation between the DHR coordinate and the
Boyer--Lindquist radius of the reference seed, the full scalar profile on the
rotation axis agrees with Eq.~\eqref{eq:YP_scalar} with a relative RMS
difference of $2.23\times10^{-3}$ over $u\le0.92$; the median pointwise
difference is $1.01\times10^{-3}$.  Figure~\ref{fig:YP_scalar_benchmark}
shows this comparison.

This agreement provides an external normalization and implementation check
independent of the CFA construction.  It is also consistent with the
observation in Ref.~\cite{Delsate2018ECS} that their small-$\xi$, small-spin
numerical solutions agree with the corresponding perturbative results.

We also continued selected spin branches beyond the common rectangle in order
to identify a conservative interpolation domain.  At
$\chi_K\simeq0.4204$ the branch remains robustly validated through
$\bar\alpha=0.01436$, including an independent horizon-regular spectral
constraint audit.  At the larger spin $\chi_K\simeq0.5491$, by contrast,
$\bar\alpha=0.01125$ slightly exceeds the adopted
${\cal C}_2$ threshold.  We therefore restrict the two-parameter analytical
family to the common rectangle
$0.005\le\bar\alpha\le0.010$ rather than using a spin-dependent outer
boundary of the numerical domain.

\section{Kerr-referenced continued-fraction representation}
\label{sec:cfa}

We now construct a global analytical representation of the numerical
stationary and axisymmetric dCS black holes introduced in
Sec.~\ref{sec:theory}.  We retain the Delsate--Herdeiro--Radu gauge of
Eq.~\eqref{eq:numerical_metric_ansatz} and the pseudoscalar
$\phi(r,\theta)$.  The use of a compact radial coordinate and continued
fractions follows the same general strategy that has proved effective for
numerical black-hole metrics in other higher-curvature
 theories~\cite{Konoplya2020EsGBCFA}.  In contrast with a direct fit of the full
metric, we subtract the exact Kerr geometry having the same $(r_H,\Omega_H)$,
\begin{equation}
F_i=F_i^{\rm K}+\Delta F_i,
\qquad
W=W_{\rm K}+\Delta W .
\label{eq:kerr_subtraction}
\end{equation}
This isolates the small dCS deformation and guarantees the correct Kerr limit
at vanishing coupling.

For arbitrary $r_H$ it is convenient to introduce
\begin{equation}
\rho=\frac{\sqrt{r^2-r_H^2}}{r_H},
\qquad
z=\frac{\rho}{1+\rho},
\qquad
y=1-z ,
\label{eq:compact_coordinate}
\end{equation}
so that the horizon and spatial infinity correspond to $z=0$ and $z=1$,
respectively.  We also use the dimensionless coupling
$\bar\alpha=\alpha/r_H^2$ and the dimensionless Kerr spin
$\chi_K=a_K/M_K$ of the reference geometry.

The angular structure required by the numerical solutions is accurately
captured by
\begin{align}
\Delta F_0={}&\bar\alpha^2\chi_K^2\Big[
y R_{00}(z)
+y^3R_{02}(z)P_2(\cos\theta)
\nonumber\\[-1mm]
&\hspace{31mm}+y^5R_{04}(z)P_4(\cos\theta)\Big],
\label{eq:cfa_f0}\\
\Delta F_1={}&\bar\alpha^2\chi_K^2\Big[
y R_{10}(z)
+y^3R_{12}(z)P_2(\cos\theta)
\nonumber\\[-1mm]
&\hspace{31mm}+y^5R_{14}(z)P_4(\cos\theta)\Big],
\label{eq:cfa_f1}\\
\Delta F_2={}&\Delta F_1+
\bar\alpha^2\chi_K^2\sin^2\theta
\Big[y^2D_0(z)
\nonumber\\[-1mm]
&\hspace{33mm}+y^4D_2(z)P_2(\cos\theta)\Big],
\label{eq:cfa_f2}\\
r_H\Delta W={}&\bar\alpha^2\chi_K\Big[
y^3{\cal W}_0(z)
\nonumber\\[-1mm]
&\hspace{28mm}+y^5{\cal W}_2(z)P_2(\cos\theta)\Big],
\label{eq:cfa_w}\\
\phi={}&\bar\alpha\chi_K\Big[
y^2S_1(z)P_1(\cos\theta)
\nonumber\\[-1mm]
&\hspace{24mm}+y^4S_3(z)P_3(\cos\theta)\Big].
\label{eq:cfa_phi}
\end{align}
The even multipoles of the metric and the odd multipoles of the pseudoscalar
implement the equatorial parity of the numerical solutions.  The explicit
powers of $y$ enforce the leading asymptotic falloff and strongly reduce
parameter degeneracies in the fit.

Each radial function is represented by one of the following low-order
continued-fraction forms,
\begin{align}
R^{[1]}(z)&=c_0+c_1y+c_2y^2,
\label{eq:cfa1}\\
R^{[2]}(z)&=c_0+c_1y+
y^2\frac{c_2}{1+c_3z},
\label{eq:cfa2}\\
R^{[3]}(z)&=c_0+c_1y+
y^2\frac{c_2}{1+\dfrac{c_3z}{1+c_4z}} .
\label{eq:cfa3}
\end{align}
Held-out radial validation selects CFA-3 for $R_{00}$, $R_{02}$, $R_{10}$,
$S_1$, and $S_3$; CFA-2 for $D_0$; and CFA-1 for the remaining radial
functions.  No pole occurs on the physical interval $0\le z\le1$; the
smallest sampled denominator margin of the final asymptotically constrained
wide-spin family is $1.210\times10^{-1}$.

\subsection{Coupling--spin family}
\label{sec:cfa_family}

The final two-parameter interpolation is built from twelve independently
validated numerical solutions,
\begin{align}
\bar\alpha&\in\{0.005,0.0075,0.010\},
\nonumber\\
\chi_K&\in\{0.202107,0.307646,0.420428,0.549139\}.
\label{eq:training_rectangle}
\end{align}
These points correspond to $\Omega_Hr_H=0.10,0.15,0.20,0.25$ on the
low-spin Kerr branch.  Each numerical solution satisfies our internal
acceptance criteria
\begin{align}
\Vert R_g\Vert_2&<5\times10^{-5},
\nonumber\\
{\rm RMS}(E^r{}_\theta)&<2\times10^{-5},
\nonumber\\
{\rm RMS}(E^r{}_r-E^\theta{}_\theta)&<5\times10^{-5}.
\label{eq:numerical_acceptance}
\end{align}
together with a relative scalar-equation residual below $10^{-10}$.  These
thresholds are numerical validation criteria adopted in this work, rather
than physical bounds on the dCS coupling.

After extracting the explicit leading factors in
Eqs.~\eqref{eq:cfa_f0}--\eqref{eq:cfa_phi}, we tested constant, coupling-only,
spin-only, additive, and bilinear dependence of the normalized CFA
coefficients.  Model selection used only the four corners of the
$(\bar\alpha,\chi_K)$ rectangle for fitting and the remaining eight points as
held-out interpolation tests.  After this cross-validation step selected the
functional dependence and radial orders, the final asymptotically constrained
coefficients were refitted jointly to all twelve numerical solutions.  Thus the
component-level errors quoted below are reconstruction errors over the final
calibration set; the light-ring and shadow observables in
Sec.~\ref{sec:shadow_validation}, which were not fit directly, provide the
independent observable validation.  Over the domain
\begin{equation}
0.005\le\bar\alpha\le0.010,
\qquad
0.202107\le\chi_K\le0.549139
\label{eq:cfa_validity_domain}
\end{equation}
the data select a particularly simple representation: all normalized profiles
except $R_{12}$ require only a linear dependence on
\begin{equation}
t=\left(\frac{\chi_K}{\chi_{\rm ref}}\right)^2,
\qquad
\chi_{\rm ref}=0.5491393984,
\label{eq:spin_coordinate}
\end{equation}
whereas $R_{12}$ is consistent with a constant profile.  Thus, for each
spin-dependent radial function,
\begin{equation}
c_j(t)=b_j^{(0)}+b_j^{(\chi)}t,
\qquad j=0,1,2,
\label{eq:coefficient_spin_fit}
\end{equation}
while the continued-fraction denominator coefficients are shared across the
validated family.  The leading monopole terms are not fitted independently.
The DHR asymptotics require
\begin{equation}
\Delta F_1=-\Delta F_0+O(r^{-2}),
\label{eq:cfa_mass_constraint}
\end{equation}
and since $y\simeq r_H/r$ at large radius we impose
\begin{equation}
c_0^{F_{10}}(t)=-c_0^{F_{00}}(t)
\label{eq:cfa_c0_constraint}
\end{equation}
exactly in the joint fit.  This guarantees that $F_0$ and $F_1$ encode the
same ADM-mass deformation at order $1/r$.  The fitted coefficients are listed
in Table~\ref{tab:cfa_coefficients}.  The apparent absence of an additional
coupling dependence in the normalized profiles should not be interpreted as a
general property of dCS gravity: it is an empirical result restricted to
Eq.~\eqref{eq:cfa_validity_domain}.  At larger coupling, our fixed-spin
continuation already shows detectable changes in the normalized profiles.

\begin{table*}[t]
\centering
\caption{Coefficients of the asymptotically constrained wide-spin CFA family. For spin-dependent rows, $c_j(t)=b_j^{(0)}+b_j^{(\chi)}t$, where $t=(\chi_K/\chi_{\rm ref})^2$ and $\chi_{\rm ref}=0.5491393984$. For $F_{12}$, $b_j^{(\chi)}=0$. The leading monopole coefficients satisfy $c_0^{F_{10}}(t)=-c_0^{F_{00}}(t)$ exactly, enforcing the common ADM-mass asymptotics. The rational denominator coefficients are shared over the validated $(\bar\alpha,\chi_K)$ domain.}
\label{tab:cfa_coefficients}
\resizebox{\textwidth}{!}{%
\begin{tabular}{c c rrr rrr rr}
\hline\hline
$R$ & order & $b_0^{(0)}$ & $b_1^{(0)}$ & $b_2^{(0)}$ & $b_0^{(\chi)}$ & $b_1^{(\chi)}$ & $b_2^{(\chi)}$ & $c_3$ & $c_4$ \\
\hline
$F00$ & 3 & -2.715033 & 65.689837 & -62.886758 & -2.022385 & 20.047600 & -17.839381 & -0.931264 & 0.059456 \\
$F02$ & 3 & -2.958535 & 13.095413 & -9.667546 & 0.020606 & -2.204773 & 2.025752 & -0.511526 & 2.015569 \\
$F04$ & 1 & 0.133195 & -0.436530 & 0.314241 & -0.445232 & 0.454306 & -0.062257 & -- & -- \\
$F10$ & 3 & 2.715033 & -19.519813 & 16.815801 & 2.022385 & -16.575652 & 14.518502 & -0.884015 & 0.080726 \\
$F12$ & 1 & -7.030812 & 15.371707 & -8.369561 & 0.000000 & 0.000000 & 0.000000 & -- & -- \\
$F14$ & 1 & -9.281316 & 19.306524 & -10.040339 & 4.855070 & -10.304009 & 5.457642 & -- & -- \\
$D0$ & 2 & -11.077771 & 15.822669 & -4.487297 & 4.440140 & -6.457949 & 1.893458 & 2.243530 & -- \\
$D2$ & 1 & -52.957152 & 1.09908e+02 & -57.395967 & 29.503716 & -63.387554 & 33.975273 & -- & -- \\
$W0$ & 1 & -12.566630 & 26.057930 & -13.491014 & 4.602676 & -8.926714 & 4.303697 & -- & -- \\
$W2$ & 1 & 0.510881 & -1.023464 & 0.512349 & 2.492345 & -5.348983 & 2.861536 & -- & -- \\
$S1$ & 3 & -2.610760 & 18.780270 & -14.328537 & 1.497761 & -6.343468 & 4.391921 & -0.528447 & 3.140849 \\
$S3$ & 3 & 0.154924 & -0.702135 & 0.541881 & 1.615000 & -9.935418 & 8.205744 & -0.736755 & 0.403249 \\
\hline\hline
\end{tabular}}
\end{table*}

For clarity, we distinguish the relative RMS error of the total metric
function,
\begin{equation}
\epsilon_X^{\rm tot}
=\frac{\|X_{\rm CFA}-X_{\rm num}\|_2}{\|X_{\rm num}\|_2},
\qquad X\in\{F_0,F_1,F_2,W\},
\label{eq:total_metric_error_definition}
\end{equation}
from an error normalized to the dCS deformation itself,
\begin{equation}
\epsilon_X^{\Delta}
=\frac{\|\Delta X_{\rm CFA}-\Delta X_{\rm num}\|_2}
{\|\Delta X_{\rm num}\|_2},
\qquad
\Delta X=X-X_{\rm K}.
\label{eq:deformation_error_definition}
\end{equation}
The worst total-metric errors over all twelve calibration points are
\begin{align}
\epsilon_{F_0}^{\rm tot}&=2.17\times10^{-5},
& \epsilon_{F_1}^{\rm tot}&=2.59\times10^{-5},
\nonumber\\
\epsilon_{F_2}^{\rm tot}&=3.66\times10^{-5},
& \epsilon_W^{\rm tot}&=1.16\times10^{-5}.
\label{eq:metric_errors}
\end{align}
The same fits, when normalized to the much smaller beyond-Kerr deformation,
have worst calibration-set errors
\begin{align}
\max\epsilon_{F_0}^{\Delta}&=0.215,
&\max\epsilon_{F_1}^{\Delta}&=0.619,
\nonumber\\
\max\epsilon_{F_2}^{\Delta}&=0.587,
&\max\epsilon_W^{\Delta}&=0.157.
\label{eq:deformation_errors_calibration}
\end{align}
The pseudoscalar reconstruction remains at the percent level, with a
worst-case relative RMS error of $1.30\%$.  The larger values in
Eq.~\eqref{eq:deformation_errors_calibration} do not contradict the much
smaller total-metric errors: throughout the present near-GR domain the dCS
metric deformation is an $O(\bar\alpha^2)$ correction to a dominant Kerr
background.  They instead quantify the more demanding question of how
faithfully the CFA reproduces the small gauge-dependent deformation itself.
This distinction is important when the parametrization is used to isolate a
beyond-Kerr signal rather than to compute geodesics in the total metric.

\subsection{Completely off-grid interpolation tests}
\label{sec:offgrid_validation}

A final interpolation test was performed after the CFA functional form,
radial orders, asymptotic constraint, and all coefficients had been frozen.
We generated two additional numerical black holes at parameter values for
which neither the coupling nor the horizon angular velocity belongs to the
calibration grid,
\begin{align}
{\rm A}:\quad &(\bar\alpha,\Omega_Hr_H,\chi_K)
=(0.00625,0.175,0.362795),
\nonumber\\
{\rm B}:\quad &(\bar\alpha,\Omega_Hr_H,\chi_K)
=(0.00875,0.225,0.481761).
\label{eq:offgrid_points}
\end{align}
Neither solution was subsequently used to alter the CFA.  On the same
$24\times16$ production grid, point A gives
\begin{align}
\|R_g\|_2&=2.39\times10^{-6},
& {\rm RMS}{\cal C}_1&=3.19\times10^{-6},
\nonumber\\
{\rm RMS}{\cal C}_2&=1.52\times10^{-5},
& R_\phi^{\rm rel}&=3.02\times10^{-14},
\label{eq:offgrid_A_residuals}
\end{align}
while point B gives
\begin{align}
\|R_g\|_2&=6.59\times10^{-7},
& {\rm RMS}{\cal C}_1&=6.85\times10^{-6},
\nonumber\\
{\rm RMS}{\cal C}_2&=3.27\times10^{-5},
& R_\phi^{\rm rel}&=2.82\times10^{-14}.
\label{eq:offgrid_B_residuals}
\end{align}
Both therefore satisfy Eqs.~\eqref{eq:numerical_acceptance} and the scalar
residual criterion.

The largest total-metric RMS error is $1.14\times10^{-5}$ at point A and
$1.71\times10^{-5}$ at point B.  At the deformation level, the errors for
$(F_0,F_1,F_2,W)$ are respectively
\begin{align}
{\rm A}:\quad &(0.116,0.361,0.441,0.140),
\nonumber\\
{\rm B}:\quad &(0.118,0.311,0.428,0.129),
\label{eq:offgrid_deformation_errors}
\end{align}
while the pseudoscalar errors are $1.09\%$ and $1.03\%$.  Thus the two unseen
points reproduce the same hierarchy seen on the calibration set: the total
metric is much more accurate than the correction normalized to its own small
amplitude.

To compare the interpolation error directly with a physical dCS signal, for a
light-ring observable $Q$ we define
\begin{equation}
\eta_Q
=\frac{|Q_{\rm CFA}-Q_{\rm num}|}
{|Q_{\rm num}-Q_{\rm K}|}.
\label{eq:signal_level_metric}
\end{equation}
For the two off-grid points the absolute relative errors in the prograde
critical impact parameter are $8.12\times10^{-7}$ and $8.81\times10^{-7}$,
and those in the retrograde impact parameter are $3.51\times10^{-6}$ and
$7.64\times10^{-6}$.  These correspond to
$\eta_{b_{\rm pro}}=0.103,0.060$ and
$\eta_{b_{\rm retro}}=0.289,0.249$ for A and B, respectively.  For the
horizontal half-width the corresponding values are
$\eta_{R_h}=0.661$ and $0.397$ because part of the dCS shift cancels between
the two shadow edges.  Table~\ref{tab:offgrid_validation} summarizes the
independent tests.

\begin{table*}[t]
\centering
\scriptsize
\setlength{\tabcolsep}{3.7pt}
\caption{Completely off-grid interpolation tests.  The two numerical solutions
were generated only after the final CFA coefficients and radial orders had
been frozen and were not used in either model selection or refitting.
$\epsilon_X^{\rm tot}$ and $\epsilon_X^{\Delta}$ denote total-metric and
deformation-normalized RMS errors, respectively, while
$\eta_Q=|Q_{\rm CFA}-Q_{\rm num}|/|Q_{\rm num}-Q_{\rm K}|$ compares the CFA
error directly with the dCS--Kerr signal in the light-ring observable $Q$.}
\label{tab:offgrid_validation}
\begin{tabular}{c c c c c c c c c c c c}
\hline\hline
point & $\bar\alpha$ & $\Omega_H r_H$ & $\chi_K$ &
$\max\epsilon_X^{\rm tot}$ & $\epsilon_{F_0}^{\Delta}$ &
$\epsilon_{F_1}^{\Delta}$ & $\epsilon_{F_2}^{\Delta}$ &
$\epsilon_W^{\Delta}$ & $\eta_{b_{\rm pro}}$ &
$\eta_{b_{\rm retro}}$ & $\eta_{R_h}$ \\
\hline
A & 0.00625 & 0.175 & 0.362795 & $1.14\times10^{-5}$ & 0.116 & 0.361 & 0.441 & 0.140 & 0.103 & 0.289 & 0.661 \\
B & 0.00875 & 0.225 & 0.481761 & $1.71\times10^{-5}$ & 0.118 & 0.311 & 0.428 & 0.129 & 0.060 & 0.249 & 0.397 \\
\hline\hline
\end{tabular}
\end{table*}

\section{Observable validation: light rings and black-hole shadows}
\label{sec:shadow_validation}

A comparison of coordinate-dependent metric components is not by itself a
sufficient accuracy test.  Continued-fraction parametrizations are more
meaningfully assessed through geodesic observables; this strategy was used,
for example, to validate analytical EsGB black-hole metrics with
shadows~\cite{Konoplya2020EsGBCFA}.  Black-hole shadows are particularly useful
in Chern--Simons gravity because photon trajectories probe the spacetime metric
directly~\cite{Okounkova2019dCSShadows}.  We therefore compare the numerical
dCS geometry and the CFA representation at the level of null geodesics.

\subsection{Equatorial light rings}

On the equatorial plane define $b=L/E$ and
\begin{equation}
{\cal H}(r,b)
=g_{\varphi\varphi}+2b\,g_{t\varphi}+b^2g_{tt}.
\label{eq:light_ring_H}
\end{equation}
The prograde and retrograde circular null orbits satisfy
\begin{equation}
{\cal H}=0,
\qquad
\partial_r{\cal H}=0.
\label{eq:light_ring_conditions}
\end{equation}
Across the twelve-point calibration rectangle, the maximum CFA errors in the
critical impact parameters are
\begin{equation}
\epsilon(b_{\rm pro})_{\max}=1.63\times10^{-6},
\qquad
\epsilon(b_{\rm retro})_{\max}=9.17\times10^{-6}.
\label{eq:impact_errors}
\end{equation}
The corresponding angular-frequency errors are the same to the quoted
precision.  The equatorial horizontal shadow half-width,
\begin{equation}
R_h=\frac{b_{\rm pro}-b_{\rm retro}}{2},
\label{eq:Rh_equatorial}
\end{equation}
is reproduced with a maximum relative error $5.87\times10^{-6}$ in this
light-ring test.  The signal-level metric in
Eq.~\eqref{eq:signal_level_metric} gives
$\eta_{b_{\rm pro}}\le0.145$ and
$\eta_{b_{\rm retro}}\le0.649$ over the twelve calibration points.  The
half-width is a less uniform signal-level diagnostic: at the weakest dCS
signals its Kerr-to-dCS shift is as small as $8.15\times10^{-7}$ in relative
terms, so cancellation between the prograde and retrograde edge shifts can
produce $\eta_{R_h}>1$ (with a maximum of $2.52$ over the calibration set)
even though the absolute CFA error in $R_h$ remains at the few-parts-in-$10^6$
level.  We therefore use the two critical impact
parameters, rather than their cancellation-sensitive combination alone, when
judging whether the CFA resolves the dCS light-ring signal.

\subsection{Full two-dimensional ray tracing}

We do not assume the existence of a Carter constant or Hamilton--Jacobi
separability.  This conservative choice is also motivated by the known loss of
the Kerr separability structure in the quadratic-in-spin dCS metric
~\cite{YagiYunesTanaka2012dCS}.  The full shadow is therefore obtained from
direct four-dimensional Hamiltonian ray tracing.  We
integrate the null Hamiltonian system
\begin{equation}
H=\frac{1}{2}g^{\mu\nu}p_\mu p_\nu=0,
\qquad
p_t=-E,\qquad p_\varphi=L,
\label{eq:null_hamiltonian}
\end{equation}
for the dynamical variables $(r,\theta,p_r,p_\theta)$.  Rays are launched from
a static orthonormal tetrad on the observer screen and are classified as
captured or escaping.  The shadow boundary is then obtained by bisection in
the screen radius at fixed polar screen angle.

For the most demanding validated point,
\begin{equation}
\bar\alpha=0.010,\qquad
\chi_K=0.549139,
\label{eq:shadow_benchmark}
\end{equation}
we computed shadows for inclinations $i=90^\circ$, $60^\circ$, and
$30^\circ$.  We characterize each contour by the area-equivalent radius
\begin{equation}
R_A=\sqrt{\frac{A}{\pi}},
\label{eq:area_radius}
\end{equation}
the horizontal and vertical half-widths $R_h$ and $R_v$, the oblateness
${\cal O}=R_v/R_h$, and a circularity measure
\begin{equation}
{\cal C}
=
\frac{\sqrt{\left\langle
(R-\langle R\rangle)^2
\right\rangle}}{\langle R\rangle},
\label{eq:circularity}
\end{equation}
where $R$ is measured from the shadow centroid.

\begin{table}[t]
\centering
\caption{CFA--numerical differences for the corrected full-shadow calculation
at $\bar\alpha=0.010$ and $\chi_K=0.549139$.  The contour error is the RMS
difference of the screen radius at common polar angles.}
\label{tab:shadow_validation}
\begin{tabular}{c c c c c}
\hline\hline
$i$ & $\epsilon(R_A)$ & $\epsilon(R_h)$ &
$\epsilon(R_v)$ & $\epsilon_\rho^{\rm RMS}$\\
\hline
$90^\circ$ & $2.96\times10^{-5}$ & $2.59\times10^{-6}$ &
$5.36\times10^{-5}$ & $3.68\times10^{-5}$\\
$60^\circ$ & $1.27\times10^{-5}$ & $2.54\times10^{-6}$ &
$1.75\times10^{-5}$ & $4.14\times10^{-5}$\\
$30^\circ$ & $2.95\times10^{-5}$ & $9.07\times10^{-6}$ &
$3.81\times10^{-5}$ & $4.01\times10^{-5}$\\
\hline\hline
\end{tabular}
\end{table}

\begin{figure*}[t]
\centering
\includegraphics[width=0.98\textwidth]{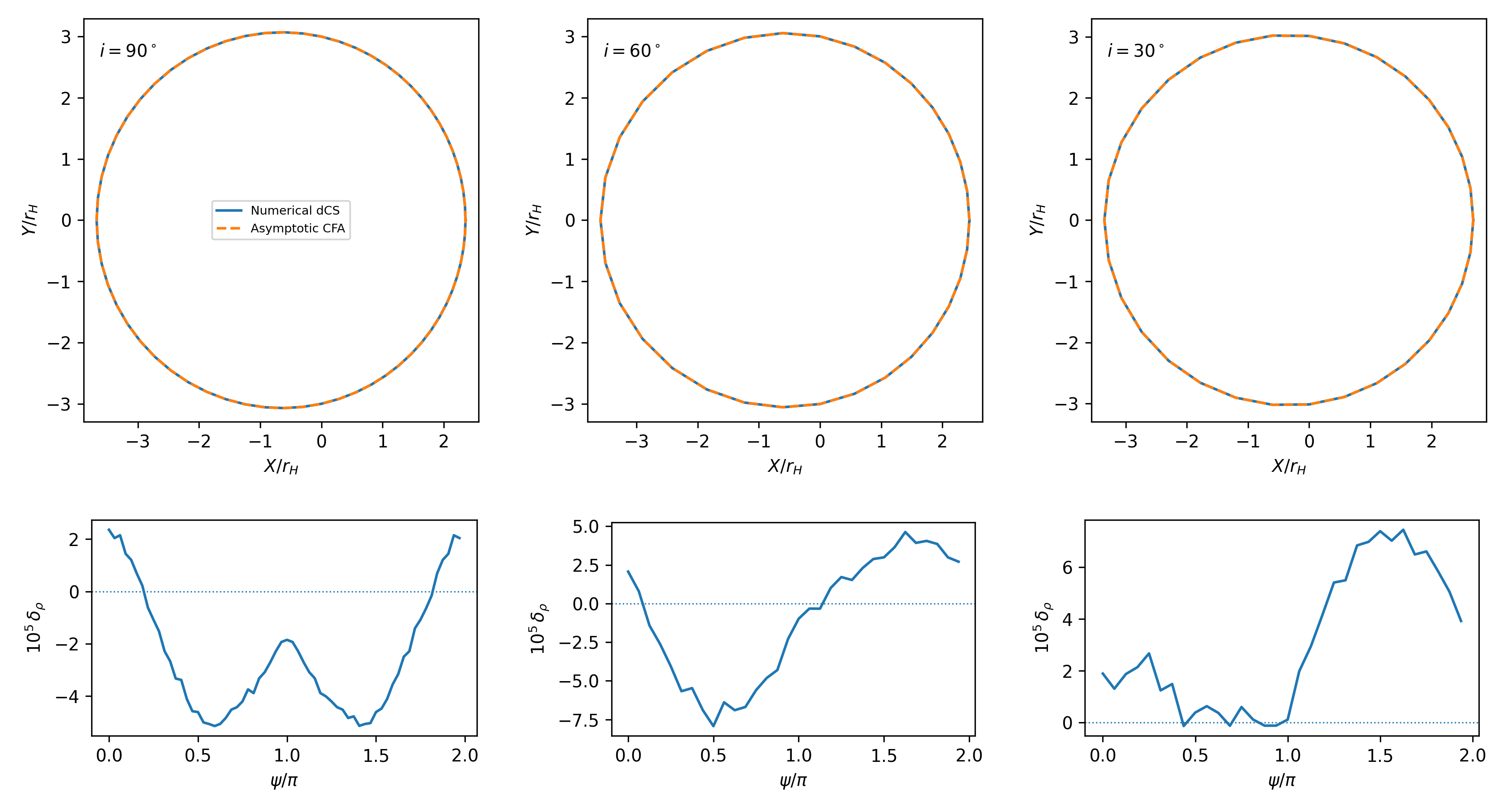}
\caption{Numerical dCS and asymptotically constrained CFA shadow comparison
for $\bar\alpha=0.010$ and $\chi_K=0.549139$ at observer inclinations
$i=90^\circ$, $60^\circ$, and $30^\circ$.  The upper row shows the shadow
contours.  Their differences are too small to resolve reliably by eye on that
scale, so the lower row shows $10^5\delta_\rho$ as a function of the screen
polar angle, with
$\delta_\rho=(\rho_{\rm CFA}-\rho_{\rm num})/\rho_{\rm num}$.}
\label{fig:shadow_contours}
\end{figure*}

The full-contour agreement remains at a few parts in $10^5$ for all three
inclinations.  At the scale of the contour plots this agreement makes the
numerical and CFA curves visually almost indistinguishable.  We therefore
show the relative contour residual explicitly in the lower row of
Fig.~\ref{fig:shadow_contours}.  The corresponding oblateness errors remain
below $6\times10^{-5}$.  Because the circularity itself is only of order
$10^{-3}$--$10^{-2}$, its \emph{relative} CFA error is larger,
approximately $0.13\%$--$0.42\%$, while the absolute shape mismatch remains
small.

\subsection{Ray-tracing convergence and asymptotic error budget}
\label{sec:raytracing_error}

We separately tested the numerical accuracy of the ray tracer.  Increasing the
number of screen directions from $32$ to $64$ at the high-spin benchmark
changes the common-angle numerical contour by only
$4.12\times10^{-6}$ in relative RMS, and the corresponding asymptotically
constrained CFA contour by $1.56\times10^{-6}$.  At $64$ directions the
numerical--CFA contour difference is $3.58\times10^{-5}$ in relative RMS, so the quoted CFA mismatch
is resolved above the ray-tracing discretization floor.

The compact numerical grid has its last finite radial point near
$r\simeq22r_H$.  To assess the influence of placing an observer farther out,
we also constructed a numerical-only asymptotic extension anchored to the
ADM charges,
\begin{align}
F_0&=\frac{c_t}{r}+O(r^{-2}),
&
F_1=F_2&=-\frac{c_t}{r}+O(r^{-2}),
\nonumber\\
W&=\frac{2J}{r^3}+O(r^{-4}).
\label{eq:asymptotic_extension}
\end{align}
Replacing the direct outer interpolation by this extension changes the
\emph{centered} shadow shape by only $6.96\times10^{-6}$ in relative RMS at
$r_{\rm obs}=50r_H$.  The absolute horizontal centroid is more sensitive to
the outer extension.  We therefore regard the shadow size, oblateness, and
centered contour as the robust validation observables, while quoting absolute
centroid shifts only with an explicit asymptotic-resolution caveat.

Taken together, the component-level tests,
equatorial light rings, and two-dimensional shadows obtained without assuming separability demonstrate
that the CFA family reproduces the numerical dCS geometry to an accuracy
sufficient for geodesic calculations throughout
Eq.~\eqref{eq:cfa_validity_domain}.  Extrapolation beyond this rectangle,
either in coupling or spin, is not used in the present analysis.

\section{Conclusions}
\label{sec:conclusions}

We have constructed a compact Kerr-referenced analytical representation of
stationary and axisymmetric black holes in dynamical Chern--Simons gravity.
For each numerical solution, the exact Kerr geometry with the same
$(r_H,\Omega_H)$ is subtracted and only the Chern--Simons deformation is
parametrized with a compact radial coordinate, low-order Legendre multipoles,
and continued fractions.  The final family covers the validated near-GR domain
$0.005\leq\bar\alpha\leq0.010$,
and 
$0.202107\leq\chi_K\leq0.549139$.
Across the twelve-point calibration set, the worst relative RMS errors of the
total metric functions are $2.17\times10^{-5}$, $2.59\times10^{-5}$,
$3.66\times10^{-5}$, and $1.16\times10^{-5}$ for
$F_0$, $F_1$, $F_2$, and $W$, respectively, while the pseudoscalar is
reconstructed at the percent level.  The continued fractions remain pole free,
and the leading $1/r$ coefficients of $F_0$ and $F_1$ are constrained to
encode a single ADM-mass deformation.

The numerical construction is supported by two independent checks.  A
$24\times16\rightarrow32\times20\rightarrow40\times24$ resolution study shows
that $M$, $J$, and the equatorial shadow half-width are stable at the
$10^{-5}$ level, while the asymptotic scalar dipole converges more slowly at
the $10^{-3}$ level.  In the slow-spin regime, the numerical scalar solution
also reproduces the Yunes--Pretorius perturbative benchmark, with
$q/(\alpha\chi_K)=0.624000$ compared with the analytical value $5/8$ and a
full-profile relative RMS difference of $2.23\times10^{-3}$.

A more stringent interpolation test was performed with two numerical solutions
generated only after the CFA coefficients had been frozen.  Their largest
total-metric RMS errors are $1.14\times10^{-5}$ and $1.71\times10^{-5}$.
Errors normalized to the much smaller dCS deformation are necessarily larger,
showing explicitly that high accuracy in the total near-Kerr metric does not
imply percent-level reconstruction of every gauge-dependent correction.
Nevertheless, for the off-grid prograde and retrograde critical impact
parameters, the CFA error remains below $29\%$ of the actual dCS--Kerr edge
shift.

The observable tests are correspondingly stronger than a component-wise metric
comparison.  Across the calibration grid, the maximum errors in the critical
impact parameters remain below $9.2\times10^{-6}$, and the horizontal
light-ring shadow half-width below $5.9\times10^{-6}$.  At
$\bar\alpha=0.010$ and $\chi_K=0.549139$, full two-dimensional ray tracing
without a separability assumption reproduces the numerical shadow size and
centered contour to a few parts in $10^5$ for inclinations from
$30^\circ$ to $90^\circ$.

The stated domain is an interpolation range, not a physical bound on dCS
gravity, and the coefficients are not extrapolated beyond it.  Signal-level
errors can become large for cancellation-sensitive observables when the
physical dCS--Kerr shift is itself extremely small.  The effective-field-theory
interpretation of dCS gravity remains distinct from the nonperturbative
boundary-value construction used here, and the present moderate-spin fit would
require a new calibration before extension toward near-extremal rotation.
Absolute shadow-centroid shifts are also more sensitive to the finite outer
numerical domain than centered shape observables.

The resulting CFA family complements theory-agnostic deformed-Kerr
parametrizations and perturbative dCS metrics
~\cite{JohannsenPsaltis2011,Johannsen2013,KonoplyaRezzollaZhidenko2016,
YunesPretorius2009dCS,YagiYunesTanaka2012dCS,WagleLiChen2024} by providing an
analytical metric calibrated directly to numerical stationary dCS solutions.
It can be used in repeated geodesic, lensing, accretion, and perturbative
calculations without interpolating the original two-dimensional numerical
data.  Natural extensions include ISCO and epicyclic-frequency systematics,
thin-disk transfer functions, perturbations on the CFA background, and a
higher-spin calibration.

\begin{acknowledgments}
S.M. gratefully acknowledges support from Grant FL--10425067111 of the Agency of Innovative Developments of Uzbekistan.
\end{acknowledgments}

\section*{Data Availability}
No external experimental or observational data were used in this study. All reported results were obtained from the equations and numerical procedures presented in the paper. The numerical values underlying the figures and the custom verification code are available from the corresponding author upon reasonable request.

\bibliographystyle{apsrev4-2}
\bibliography{ecs_core_refs}

@article{Delsate2018ECS,
  author  = {Delsate, T{\'e}rence and Herdeiro, Carlos and Radu, Eugen},
  title   = {Non-perturbative spinning black holes in dynamical Chern--Simons gravity},
  journal = {Physics Letters B},
  volume  = {787},
  pages   = {8--15},
  year    = {2018},
  doi     = {10.1016/j.physletb.2018.09.060}
}

@article{Konoplya2020EsGBCFA,
  author  = {Konoplya, R. A. and Pappas, Thomas D. and Zhidenko, A.},
  title   = {Einstein-scalar--Gauss--Bonnet black holes: Analytical approximation for the metric and applications to calculations of shadows},
  journal = {Physical Review D},
  volume  = {101},
  number  = {4},
  pages   = {044054},
  year    = {2020},
  doi     = {10.1103/PhysRevD.101.044054}
}

@article{Okounkova2019dCSShadows,
  author  = {Okounkova, Maria and Scheel, Mark and Teukolsky, Saul A.},
  title   = {Numerical black hole initial data and shadows in dynamical Chern--Simons gravity},
  journal = {Classical and Quantum Gravity},
  volume  = {36},
  number  = {5},
  pages   = {054001},
  year    = {2019},
  doi     = {10.1088/1361-6382/aafcdf}
}

@article{YunesPretorius2009dCS,
  author  = {Yunes, Nicol{\'a}s and Pretorius, Frans},
  title   = {Dynamical Chern--Simons modified gravity: Spinning black holes in the slow-rotation approximation},
  journal = {Physical Review D},
  volume  = {79},
  number  = {8},
  pages   = {084043},
  year    = {2009},
  doi     = {10.1103/PhysRevD.79.084043}
}

@article{YagiYunesTanaka2012dCS,
  author  = {Yagi, Kent and Yunes, Nicol{\'a}s and Tanaka, Takahiro},
  title   = {Slowly rotating black holes in dynamical Chern--Simons gravity: Deformation quadratic in the spin},
  journal = {Physical Review D},
  volume  = {86},
  number  = {4},
  pages   = {044037},
  year    = {2012},
  doi     = {10.1103/PhysRevD.86.044037},
  note    = {Erratum: Phys. Rev. D 89, 049902 (2014)}
}

@article{AliHaimoudChen2011dCS,
  author  = {Ali-Ha{\"i}moud, Yacine and Chen, Yanbei},
  title   = {Slowly rotating stars and black holes in dynamical Chern--Simons gravity},
  journal = {Physical Review D},
  volume  = {84},
  number  = {12},
  pages   = {124033},
  year    = {2011},
  doi     = {10.1103/PhysRevD.84.124033}
}

@article{DelsateHilditchWitek2015,
  author  = {Delsate, T{\'e}rence and Hilditch, David and Witek, Helvi},
  title   = {Initial value formulation of dynamical Chern--Simons gravity},
  journal = {Physical Review D},
  volume  = {91},
  number  = {2},
  pages   = {024027},
  year    = {2015},
  doi     = {10.1103/PhysRevD.91.024027}
}

@article{KonoplyaRezzollaZhidenko2016,
  author  = {Konoplya, R. A. and Rezzolla, Luciano and Zhidenko, A.},
  title   = {General parametrization of axisymmetric black holes in metric theories of gravity},
  journal = {Physical Review D},
  volume  = {93},
  number  = {6},
  pages   = {064015},
  year    = {2016},
  doi     = {10.1103/PhysRevD.93.064015}
}

@article{JackiwPi2003,
  author  = {Jackiw, R. and Pi, S.-Y.},
  title   = {Chern--Simons modification of general relativity},
  journal = {Physical Review D},
  volume  = {68},
  pages   = {104012},
  year    = {2003},
  doi     = {10.1103/PhysRevD.68.104012}
}

@article{AlexanderYunes2009Review,
  author  = {Alexander, Stephon and Yunes, Nicol{\'a}s},
  title   = {Chern--Simons modified general relativity},
  journal = {Physics Reports},
  volume  = {480},
  pages   = {1--55},
  year    = {2009},
  doi     = {10.1016/j.physrep.2009.07.002}
}

@article{KonnoMatsuyamaTanda2007,
  author  = {Konno, Kohkichi and Matsuyama, Toyoki and Tanda, Satoshi},
  title   = {Does a black hole rotate in Chern--Simons modified gravity?},
  journal = {Physical Review D},
  volume  = {76},
  pages   = {024009},
  year    = {2007},
  doi     = {10.1103/PhysRevD.76.024009}
}

@article{KonnoMatsuyamaTanda2009,
  author  = {Konno, Kohkichi and Matsuyama, Toyoki and Tanda, Satoshi},
  title   = {Rotating Black Hole in Extended Chern--Simons Modified Gravity},
  journal = {Progress of Theoretical Physics},
  volume  = {122},
  pages   = {561--568},
  year    = {2009},
  doi     = {10.1143/PTP.122.561}
}

@article{CardosoGualtieri2009,
  author  = {Cardoso, V{\'i}tor and Gualtieri, Leonardo},
  title   = {Perturbations of Schwarzschild black holes in dynamical Chern--Simons modified gravity},
  journal = {Physical Review D},
  volume  = {80},
  pages   = {064008},
  year    = {2009},
  doi     = {10.1103/PhysRevD.80.064008},
  note    = {Erratum: Phys. Rev. D 81, 089903 (2010)}
}

@article{MolinaPaniCardosoGualtieri2010,
  author  = {Molina, C. and Pani, Paolo and Cardoso, V{\'i}tor and Gualtieri, Leonardo},
  title   = {Gravitational signature of Schwarzschild black holes in dynamical Chern--Simons gravity},
  journal = {Physical Review D},
  volume  = {81},
  pages   = {124021},
  year    = {2010},
  doi     = {10.1103/PhysRevD.81.124021}
}

@article{GarfinklePretoriusYunes2010,
  author  = {Garfinkle, David and Pretorius, Frans and Yunes, Nicol{\'a}s},
  title   = {Linear stability analysis and the speed of gravitational waves in dynamical Chern--Simons modified gravity},
  journal = {Physical Review D},
  volume  = {82},
  pages   = {041501},
  year    = {2010},
  doi     = {10.1103/PhysRevD.82.041501}
}

@article{AmarillaEiroaGiribet2010,
  author  = {Amarilla, Leonardo D. and Eiroa, Ernesto F. and Giribet, Gast{\'o}n},
  title   = {Null geodesics and shadow of a rotating black hole in extended Chern--Simons modified gravity},
  journal = {Physical Review D},
  volume  = {81},
  pages   = {124045},
  year    = {2010},
  doi     = {10.1103/PhysRevD.81.124045}
}

@article{ChenJing2010,
  author  = {Chen, Songbai and Jing, Jiliang},
  title   = {Geodetic precession and strong gravitational lensing in dynamical Chern--Simons-modified gravity},
  journal = {Classical and Quantum Gravity},
  volume  = {27},
  pages   = {225006},
  year    = {2010},
  doi     = {10.1088/0264-9381/27/22/225006}
}

@article{HarkoKovacsLobo2010,
  author  = {Harko, Tiberiu and Kov{\'a}cs, Zolt{\'a}n and Lobo, Francisco S. N.},
  title   = {Thin accretion disk signatures in dynamical Chern--Simons-modified gravity},
  journal = {Classical and Quantum Gravity},
  volume  = {27},
  pages   = {105010},
  year    = {2010},
  doi     = {10.1088/0264-9381/27/10/105010}
}

@article{MotohashiSuyama2012,
  author  = {Motohashi, Hayato and Suyama, Teruaki},
  title   = {Black hole perturbation in nondynamical and dynamical Chern--Simons gravity},
  journal = {Physical Review D},
  volume  = {85},
  pages   = {044054},
  year    = {2012},
  doi     = {10.1103/PhysRevD.85.044054}
}

@article{AdakDereli2012,
  author  = {Adak, Muzaffer and Dereli, Tekin},
  title   = {String-inspired Chern--Simons modified gravity in four dimensions},
  journal = {The European Physical Journal C},
  volume  = {72},
  pages   = {1979},
  year    = {2012},
  doi     = {10.1140/epjc/s10052-012-1979-0}
}

@article{KonnoTakahashi2014,
  author  = {Konno, Kohkichi and Takahashi, Rohta},
  title   = {Scalar field excited around a rapidly rotating black hole in Chern--Simons modified gravity},
  journal = {Physical Review D},
  volume  = {90},
  pages   = {064011},
  year    = {2014},
  doi     = {10.1103/PhysRevD.90.064011}
}

@article{McNeesSteinYunes2016,
  author  = {McNees, Robert and Stein, Leo C. and Yunes, Nicol{\'a}s},
  title   = {Extremal black holes in dynamical Chern--Simons gravity},
  journal = {Classical and Quantum Gravity},
  volume  = {33},
  pages   = {235013},
  year    = {2016},
  doi     = {10.1088/0264-9381/33/23/235013}
}

@article{OkounkovaSteinScheel2017,
  author  = {Okounkova, Maria and Stein, Leo C. and Scheel, Mark A. and Teukolsky, Saul A.},
  title   = {Numerical binary black hole mergers in dynamical Chern--Simons gravity: Scalar field},
  journal = {Physical Review D},
  volume  = {96},
  pages   = {044020},
  year    = {2017},
  doi     = {10.1103/PhysRevD.96.044020}
}

@article{OkounkovaSteinMoxon2020,
  author  = {Okounkova, Maria and Stein, Leo C. and Moxon, Jordan and Scheel, Mark A.},
  title   = {Numerical relativity simulation of GW150914 beyond general relativity},
  journal = {Physical Review D},
  volume  = {101},
  pages   = {104016},
  year    = {2020},
  doi     = {10.1103/PhysRevD.101.104016}
}

@article{SrivastavaChenShankaranarayanan2021,
  author  = {Srivastava, Manu and Chen, Yanbei and Shankaranarayanan, S.},
  title   = {Analytical computation of quasinormal modes of slowly rotating black holes in dynamical Chern--Simons gravity},
  journal = {Physical Review D},
  volume  = {104},
  pages   = {064034},
  year    = {2021},
  doi     = {10.1103/PhysRevD.104.064034}
}

@article{DonevaYazadjiev2021,
  author  = {Doneva, Daniela D. and Yazadjiev, Stoytcho S.},
  title   = {Spontaneously scalarized black holes in dynamical Chern--Simons gravity: Dynamics and equilibrium solutions},
  journal = {Physical Review D},
  volume  = {103},
  pages   = {083007},
  year    = {2021},
  doi     = {10.1103/PhysRevD.103.083007}
}

@article{WagleYunesSilva2022,
  author  = {Wagle, Pratik and Yunes, Nicol{\'a}s and Silva, Hector O.},
  title   = {Quasinormal modes of slowly-rotating black holes in dynamical Chern--Simons gravity},
  journal = {Physical Review D},
  volume  = {105},
  pages   = {124003},
  year    = {2022},
  doi     = {10.1103/PhysRevD.105.124003}
}

@article{AlexanderGabadadzeJenks2023,
  author  = {Alexander, Stephon and Gabadadze, Gregory and Jenks, Leah},
  title   = {Black hole superradiance in dynamical Chern--Simons gravity},
  journal = {Physical Review D},
  volume  = {107},
  pages   = {084016},
  year    = {2023},
  doi     = {10.1103/PhysRevD.107.084016}
}

@article{WagleLiChen2024,
  author  = {Wagle, Pratik and Li, Dongjun and Chen, Yanbei},
  title   = {Perturbations of spinning black holes in dynamical Chern--Simons gravity: Slow rotation equations},
  journal = {Physical Review D},
  volume  = {109},
  pages   = {104029},
  year    = {2024},
  doi     = {10.1103/PhysRevD.109.104029}
}

@article{CanoRuiperez2019,
  author  = {Cano, Pablo A. and Ruip{\'e}rez, Alejandro},
  title   = {Leading higher-derivative corrections to Kerr geometry},
  journal = {Journal of High Energy Physics},
  volume  = {2019},
  pages   = {189},
  year    = {2019},
  doi     = {10.1007/JHEP05(2019)189},
  note    = {Erratum: JHEP 03 (2020) 187}
}

@article{CanoFransenHertog2022,
  author  = {Cano, Pablo A. and Fransen, Kwinten and Hertog, Thomas},
  title   = {Gravitational ringing of rotating black holes in higher-derivative gravity},
  journal = {Physical Review D},
  volume  = {105},
  pages   = {024064},
  year    = {2022},
  doi     = {10.1103/PhysRevD.105.024064}
}

@article{RezzollaZhidenko2014,
  author  = {Rezzolla, Luciano and Zhidenko, A.},
  title   = {New parametrization for spherically symmetric black holes in metric theories of gravity},
  journal = {Physical Review D},
  volume  = {90},
  pages   = {084009},
  year    = {2014},
  doi     = {10.1103/PhysRevD.90.084009}
}

@article{KonoplyaRezzollaZhidenko2016Shadow,
  author  = {Younsi, Ziri and Zhidenko, Alexander and Rezzolla, Luciano and Konoplya, Roman and Mizuno, Yosuke},
  title   = {New method for shadow calculations: Application to parametrized axisymmetric black holes},
  journal = {Physical Review D},
  volume  = {94},
  pages   = {084025},
  year    = {2016},
  doi     = {10.1103/PhysRevD.94.084025}
}

@article{JohannsenPsaltis2011,
  author  = {Johannsen, Tim and Psaltis, Dimitrios},
  title   = {Metric for rapidly spinning black holes suitable for strong-field tests of the no-hair theorem},
  journal = {Physical Review D},
  volume  = {83},
  pages   = {124015},
  year    = {2011},
  doi     = {10.1103/PhysRevD.83.124015}
}

@article{Johannsen2013,
  author  = {Johannsen, Tim},
  title   = {Regular black hole metric with three constants of motion},
  journal = {Physical Review D},
  volume  = {88},
  pages   = {044002},
  year    = {2013},
  doi     = {10.1103/PhysRevD.88.044002}
}

@article{VigelandHughes2010,
  author  = {Vigeland, Sarah J. and Hughes, Scott A.},
  title   = {Spacetime and orbits of bumpy black holes},
  journal = {Physical Review D},
  volume  = {81},
  pages   = {024030},
  year    = {2010},
  doi     = {10.1103/PhysRevD.81.024030}
}

@article{RezzollaMizunoYounsi2018,
  author  = {Rezzolla, Luciano and Mizuno, Yosuke and Younsi, Ziri},
  title   = {The current ability to test theories of gravity with black hole shadows},
  journal = {Nature Astronomy},
  volume  = {2},
  pages   = {585--590},
  year    = {2018},
  doi     = {10.1038/s41550-018-0449-5}
}

@article{PerlickTsupko2022,
  author  = {Perlick, Volker and Tsupko, Oleg Yu.},
  title   = {Calculating black hole shadows: Review of analytical studies},
  journal = {Physics Reports},
  volume  = {947},
  pages   = {1--39},
  year    = {2022},
  doi     = {10.1016/j.physrep.2021.10.004}
}

@article{NiJiangBambi2016,
  author  = {Ni, Yueying and Jiang, Jiachen and Bambi, Cosimo},
  title   = {Testing the Kerr metric with the iron line and the KRZ parametrization},
  journal = {Journal of Cosmology and Astroparticle Physics},
  volume  = {2016},
  number  = {09},
  pages   = {014},
  year    = {2016},
  doi     = {10.1088/1475-7516/2016/09/014}
}

@article{ZhouChenJing2021,
  author  = {Zhou, Xuan and Chen, Songbai and Jing, Jiliang},
  title   = {Chaotic motion of scalar particle coupling to Chern--Simons invariant in Kerr black hole spacetime},
  journal = {The European Physical Journal C},
  volume  = {81},
  year    = {2021},
  doi     = {10.1140/epjc/s10052-021-09042-7}
}

@article{BertiYagiYunes2018,
  author  = {Berti, Emanuele and Yagi, Kent and Yunes, Nicol{\'a}s},
  title   = {Extreme gravity tests with gravitational waves from compact binary coalescences: (I) inspiral--merger},
  journal = {General Relativity and Gravitation},
  volume  = {50},
  year    = {2018},
  doi     = {10.1007/s10714-018-2362-8}
}

@article{YunesSiemens2013,
  author  = {Yunes, Nicol{\'a}s and Siemens, Xavier},
  title   = {Gravitational-Wave Tests of General Relativity with Ground-Based Detectors and Pulsar-Timing Arrays},
  journal = {Living Reviews in Relativity},
  volume  = {16},
  year    = {2013},
  doi     = {10.12942/lrr-2013-9}
}

@article{ChamberlainYunes2017,
  author  = {Chamberlain, Katie and Yunes, Nicol{\'a}s},
  title   = {Theoretical physics implications of gravitational wave observation with future detectors},
  journal = {Physical Review D},
  volume  = {96},
  pages   = {084039},
  year    = {2017},
  doi     = {10.1103/PhysRevD.96.084039}
}

@article{Berti2019KerrTests,
  author  = {Berti, Emanuele},
  title   = {Topical collection: Testing the Kerr spacetime with gravitational-wave and electromagnetic observations},
  journal = {General Relativity and Gravitation},
  volume  = {51},
  year    = {2019},
  doi     = {10.1007/s10714-019-2622-2}
}

@article{Krawczynski2012,
  author  = {Krawczynski, Henric},
  title   = {Tests of general relativity in the strong-gravity regime based on X-ray spectropolarimetric observations of black holes in X-ray binaries},
  journal = {The Astrophysical Journal},
  volume  = {754},
  pages   = {133},
  year    = {2012},
  doi     = {10.1088/0004-637X/754/2/133}
}

@article{BertiCardosoCarullo2026,
  author  = {Berti, Emanuele and Cardoso, Vitor and Carullo, Gregorio and others},
  title   = {Black hole spectroscopy: from theory to experiment},
  journal = {Classical and Quantum Gravity},
  volume  = {43},
  pages   = {123001},
  year    = {2026},
  doi     = {10.1088/1361-6382/ae59e2}
}

@article{XieZhangSilva2021,
  author  = {Xie, Yiqi and Zhang, Jun and Silva, Hector O.},
  title   = {Square Peg in a Circular Hole: Choosing the Right Ansatz for Isolated Black Holes in Generic Gravitational Theories},
  journal = {Physical Review Letters},
  volume  = {126},
  pages   = {241104},
  year    = {2021},
  doi     = {10.1103/PhysRevLett.126.241104}
}

@article{IbadovKleihausKunzMurodov2021,
  author  = {Ibadov, Rustam and Kleihaus, Burkhard and Kunz, Jutta and Murodov, Sardor},
  title   = {Scalarized Nutty Wormholes},
  journal = {Symmetry},
  volume  = {13},
  number  = {1},
  pages   = {89},
  year    = {2021},
  doi     = {10.3390/sym13010089}
}

@article{IbadovKleihausKunzMurodov2022,
  author  = {Ibadov, Rustam and Kleihaus, Burkhard and Kunz, Jutta and Murodov, Sardor},
  title   = {Wormhole solutions with {NUT} charge in higher curvature theories},
  journal = {Arabian Journal of Mathematics},
  volume  = {11},
  number  = {1},
  pages   = {31--41},
  year    = {2022},
  doi     = {10.1007/s40065-021-00350-0}
}

@article{CunhaHerdeiroKleihausKunzRadu2017,
  author  = {Cunha, Pedro V. P. and Herdeiro, Carlos A. R. and Kleihaus, Burkhard and Kunz, Jutta and Radu, Eugen},
  title   = {Shadows of Einstein--dilaton--Gauss--Bonnet black holes},
  journal = {Physics Letters B},
  volume  = {768},
  pages   = {373--379},
  year    = {2017},
  doi     = {10.1016/j.physletb.2017.03.020}
}

@article{MurodovRayimbaevAhmedovKarimbaev2023,
  author  = {Murodov, Sardor and Rayimbaev, Javlon and Ahmedov, Bobomurat and Karimbaev, Eldor},
  title   = {Quasiperiodic Oscillations and Dynamics of Test Particles around Quasi- and Non-Schwarzschild Black Holes},
  journal = {Universe},
  volume  = {9},
  number  = {9},
  pages   = {391},
  year    = {2023},
  doi     = {10.3390/universe9090391}
}

\end{document}